\documentclass[aps,prd,twocolumn,nofootinbib]{revtex4}

\usepackage{graphicx}
\usepackage{hyperref}
\usepackage{slashed}
\usepackage{color}

\begin{document}

\title{Production of ${\bar{B}^{0}_{s}}$ or ${{B}^{0}_{s}}$ and its excited states via ${t}$-quark or ${\bar{t}}$-quark decays}

\author{Qi-Li Liao $^{1}$}
\email{xiaosueer@163.com}
\author{Yan Yu $^{1}$}
\author{Ya Deng $^{1}$}
\author{Guang-Chuan Wang $^{1}$}
\author{Jia-Wei Zhang $^{2}$}
\email{jwzhang@cqust.edu.cn}
\author{Guo-Ya Xie $^{3}$}

\address{$^{1}$ College Mobile Telecommunications Chongqing  University of Posts and Telecom, Chongqing 401520, People's Republic of China\\
$^{2}$ Department of Physics, Chongqing University of Science and Technology, Chongqing 401331, People's Republic of China\\
$^{3}$ Chongqing  University of Posts and Telecom, Chongqing 400065, People's Republic of China}

\date{\today}

\begin{abstract}
    In this work we evaluate the masses of the $|(b\bar{s})[n]\rangle$ or $|(\bar{b}s)[n]\rangle$ quarkonium ($\bar{B}^{0}_{s}$ or ${B}^{0}_{s}$ meson) under the B.T. potential, and the values of the Schr${\rm \ddot{o}}$dinger radial wave function at the origin of the $|(b\bar{s})[n]\rangle$ or $|(\bar{b}s)[n]\rangle$ quarkonium within the five potential models. Then we investigate a systematic study on the production of the $|(b\bar{s})[n]\rangle$ or $|(\bar{b}s)[n]\rangle$ quarkonium via top quark or antitop quark decays in the color-singlet QCD factorization formula (CSQCDFF), i.e., the two $S$-wave states, $|(b\bar{s})[1^1S_0] \rangle$ (or $|(\bar{b}s)[1^1S_0] \rangle$) and $|(b\bar{s})[1^3S_1] \rangle$ (or $|(\bar{b}s)[1^3S_1] \rangle$), and its four $P$-wave excited states, $|(b\bar{s})[1^1P_1] \rangle$ (or $|(\bar{b}s)[1^1P_1] \rangle$) and $|(b\bar{s})[1^3P_J] \rangle$ (or $|(\bar{b}s)[1^3P_J] \rangle$) (with $J =[0, 1, 2]$). For deriving compact analytical results for complex processes, the ``improved trace technology" is adopted to deal with the decay channels at the amplitudes. Moreover, various differential distributions and uncertainties of the concerned processes are analyzed carefully. By adding the uncertainties caused by the ${b}$ and ${s}$-quark masses in quadrature, we obtain $\Gamma{(t\to |(b\bar{s})[n]\rangle +W^{+}s)}=14.19^{+4.36}_{-3.20}$~MeV. At the LHC with the luminosity ${\cal L}\propto 10^{34}cm^{-2}s^{-1}$ and the center-of-mass energy $\sqrt{S}=14$ TeV, sizable $|(b\bar{s})[n]\rangle$ or $|(\bar{b}s)[n]\rangle$ meson events can be produced through ${t}$-quark or ${\bar{t}}$-quark decays; i.e., about $1.3~\times10^6$ ${\bar{B}^0_s}$ or ${B^0_s}$ events per year can be obtained.\\


\end{abstract}

\maketitle

\section{Introduction}

The CDF and D0 Collaborations at the Tevatron have successfully collected numerous data for the ${\bar{B}^{0}_{s}}$ and ${B^{0}_{s}}$ meson~\cite{dt, ta, aab, vma}. The LHC shall also provide a good platform to study the properties of the ${\bar{B}^{0}_{s}}$ and ${B^{0}_{s}}$ meson, e.g., recent results and review papers on the ${\bar{B}^{0}_{s}}$ and ${B^{0}_{s}}$ meson measurements and searches by the LHCb, CMS, and ATLAS Collaborations at the LHC can be found in Refs.~\cite{bs1, bs2, bs3, bs4, bs5, bs6}.

The ${\bar{B}^{0}_{s}}$ and ${B^{0}_{s}}$ meson form an interesting experiment and theory for the study of the quantum chromodynamics (QCD) and new physics (NP), which can be used to study those interesting topics as the QCD model building, the electroweak symmetry breaking mechanism, charge-parity (CP) violation, new physics beyond the Standard Model (SM) and etc.~\cite{bs16, bs17, bs18, bs19}. Taking into account the prospects of the ${\bar{B}^{0}_{s}}$ or ${B^{0}_{s}}$ meson production and decay at the TEVATRON and at the running LHC, the future numerous data require more accurate theoretical predictions on its hadronic production and indirect production through  weak-decay processes.

In general, the band states of the ${(\bar{c}{c})}$ ,${(\bar{b}{c})}$ and ${(\bar{b}{b})}$ quarkonium are the heavy quarkonium. But the ${B^{0}_{s}}$ meson contains $b$-heavy quark, which can also be approximated as the heavy quarkonium. Thus, it is beneficial to study the production and decay properties of the ${B^{0}_{s}}$ meson. For the ${\bar{B}^{0}_{s}}$ or ${B^{0}_{s}}$ meson direct production, various investigations has been carried out studied in detail in Refs.~\cite{zy, bq, Zhang1, Zhang2}. The ${\bar{B}^{0}_{s}}$ or ${B^{0}_{s}}$ meson indirect production through the $Z^0$ boson and high excited $\Upsilon({\rm nS})$ meson decays has been measured and discussed in Refs.~\cite{productionBs1,productionBs2, productionBs3, productionBs4, productionBs5, productionBs6}. Although in literature the direct hadronic production of the ${\bar{B}^{0}_{s}}$ or ${B^{0}_{s}}$ meson has been thoroughly studied in Refs.~\cite{Zhang1, Zhang2} and CDF discovered the meson which just comes from the hadroproduction.  As a compensation to understanding the production mechanisms, and also testing the perturbative quantum chromodynamics (pQCD) and the QCD factorization formula. It is quite interesting that to study the indirectly production of the ${\bar{B}^{0}_{s}}$ or ${B^{0}_{s}}$ through top quark or antitop quark decays, especially on considering the fact that numerous ${t}$-quark or ${\bar{t}}$-quark may be produced at the LHC. At the LHC running at the center-of-mass energy $\sqrt{S}=14$ TeV and luminosity raising up to ${\cal L}\propto 10^{34}cm^{-2}s^{-1}$~\cite{bc1,bc2}, about $10^{8}$ $t\bar{t}$ quark per year will be produced. More ${t}$-quark or ${\bar{t}}$-quark rare decays can be adopted for precise studies at the LHC. This paper is devoted to study the indirect production of the ${\bar{B}^{0}_{s}}$ or ${B^{0}_{s}}$ meson via ${t}$-quark or ${\bar{t}}$-quark decays. The theoretical studies of the bound states is based on the factorization formula, which includes the color-singlet QCD factorization formula (CSQCDFF) \cite{getg}, the nonrelativistic quantum chromodynamics (NRQCD) \cite{nrqcd}, the $k_{\perp}$-factorization \cite{smf} and etc. The CSQCDFF is a useful method for studying on the production and decays of the quarkinum. According to the CSQCDFF, the processes of the production and decays of the quarkinum can be factorized as two parts of the perturbatively calculable short-distance coefficients and the nonperturbative long-distance factors. In present paper it is investigated a systematic study of the indirect production of the $|(b\bar{s})[n]\rangle$ or $|(\bar{b}s)[n]\rangle$ quarkonium via $t$-quark or $\bar{t}$-quark decays within the CSQCDFF. To derive the analytical expression of the pQCD calculable conventional squared amplitudes becomes too complex and lengthy for more (massive) particles in the final states and for the excited states to be generated for the emergence of massive-fermion lines in the Feynman diagrams, especially to derive the squared amplitudes of the $P$-wave states. In Refs.~\cite{tbc2, zbc0, zbc1, zbc2, wbc1, wbc2, lyd, lx}, the ``improved trace technology'' is suggested and developed to solve the problem. Here we adopt the ``improved trace technology" to derive the analytical expression for the above-mentioned processes. Without confusing and for simplifying the statements, later on we will not distinguish the ${\bar{B}^{0}_{s}}$ and ${B^{0}_{s}}$ meson unless necessary. Because of the production of the $|(b\bar{s})[n]\rangle$ or $|(\bar{b}s)[n]\rangle$ meson through ${t}$-quark or ${\bar{t}}$-quark decays channels, i.e., $t\to |(b\bar{s})[n]\rangle+W^{+}s$ and $\bar{t}\to |(\bar{b}s)[n]\rangle+W^{-}\bar{s}$, and $t\to |(\bar{b}s)[n]\rangle+W^{+}b$ and $\bar{t}\to |(b\bar{s})[n]\rangle+W^{-}\bar{b}$, are symmetric in the interchange from particle to anti-particle.

The rest is organized as follows. In Sec.~II, we introduce the calculation techniques for the above-mentioned ${t}$-quark and ${\bar{t}}$-quark decays to the $|(b\bar{s})[n]\rangle$ quarkonium. In order to calculate the production of the $|(b\bar{s})[n]\rangle$ quarkonium and its excited states, we evaluate the masses of the $|(b\bar{s})[n]\rangle$ quarkonium with the energy eigenvalue of the nonrelativistic Schr${\rm \ddot{o}}$dinger equation. In Sec.III, we calculate and tabulate the masses of the $|(b\bar{s})[n]\rangle$ quarkonium and its values of the Schr${\rm \ddot{o}}$dinger radial wave functions at the origin. Then, we investigate a systematic study on the production of $|(b\bar{s})[n]\rangle$ through ${t}$-quark and ${\bar{t}}$-quark decays channels, i.e., $t\to |(b\bar{s})[n]\rangle+W^{+}s$ and $\bar{t} \to |(b\bar{s})[n]\rangle+W^{-}\bar{b}$, where $|(b\bar{s})[n]\rangle$ stands for $|(b\bar{s})[1^1S_1]\rangle$, $|(b\bar{s})[1^3S_1]\rangle$, $|(b\bar{s})[1^1P_1]\rangle$, and $|(b\bar{s})[1^3P_J]\rangle$ (with $J=[0,1,2]$). We make some discussions on the various differential distributions and the uncertainties of the decays widths by the masses of the $|(b\bar{s})[n]\rangle$ quarkonium and the correspondingly values of the Schr${\rm \ddot{o}}$dinger radial wave functions at the origin. The final section is reserved for a summary.

\section{Calculation techniques and matrix element $\langle{\cal O}^H(n)\rangle$}

\subsection{Calculation techniques}

For the production of the $|(b\bar{s})[n]\rangle$ meson through ${t}$-quark and ${\bar{t}}$-quark decays channels, $ t (k) \to |(b\bar{s})[n]\rangle(q_3) +W^{+}(q_2) + s(q_1)$ and $ \bar{t} (k) \to |(b\bar{s})[n]\rangle(q_3) +W^{-}(q_2) + \bar{b}(q_1)$, where $k$ and $q_i$ ($i=1, 2, 3$) are the momenta of the corresponding particles, according to the CSQCDFF, its total decay widths $d\Gamma$ can be factorized as \cite{getg}

\begin{eqnarray}
d\Gamma&=&\sum_{n} d\hat\Gamma(t\to |(b\bar{s})[n]\rangle+ W^{+}s) \langle{\cal O}^H(n) \rangle,\nonumber\\
d\Gamma&=&\sum_{n} d\hat\Gamma(\bar{t}\to |(b\bar{s})[n]\rangle+W^{-}\bar{b}) \langle{\cal O}^H(n) \rangle.
\end{eqnarray}
where the nonperturbative matrix element $\langle{\cal O}^{H}(n)\rangle$ describes the hadronization of a $b\bar{s}$ pair into the observable quark state $H$ and is proportional to the transition probability of the perturbative state $b\bar{s}$ into the bound state $|(b\bar{s})[n]\rangle$ quarkonium. The matrix elements $\langle{\cal O}^{H}(n)\rangle$ for the color-singlet components can be directly related to the Schr${\rm \ddot{o}}$dinger wave functions at the origin for the $S$-wave states and the first derivative of the wave functions at the origin for the $P$-wave states, which can be computed via the potential models~\cite{lx, pot1, pot2, pot3, pot4, pot5}. The nonperturbative matrix element $\langle{\cal O}^{H}(n)\rangle$ can also be obtained via potential NRQCD (pNRQCD)~\cite{pnrqcd1, yellow} and/or lattice QCD~\cite{lat1}.

The hard-scattering amplitudes for specified processes can be dealt with
\begin{eqnarray}
&& t\rightarrow|(b\bar{s})[n]\rangle+W^{+}s,\nonumber\\
&&\bar{t}\rightarrow|(b\bar{s})[n]\rangle+W^{-}\bar{b}.
\end{eqnarray}

\begin{figure}
\includegraphics[width=0.40\textwidth]{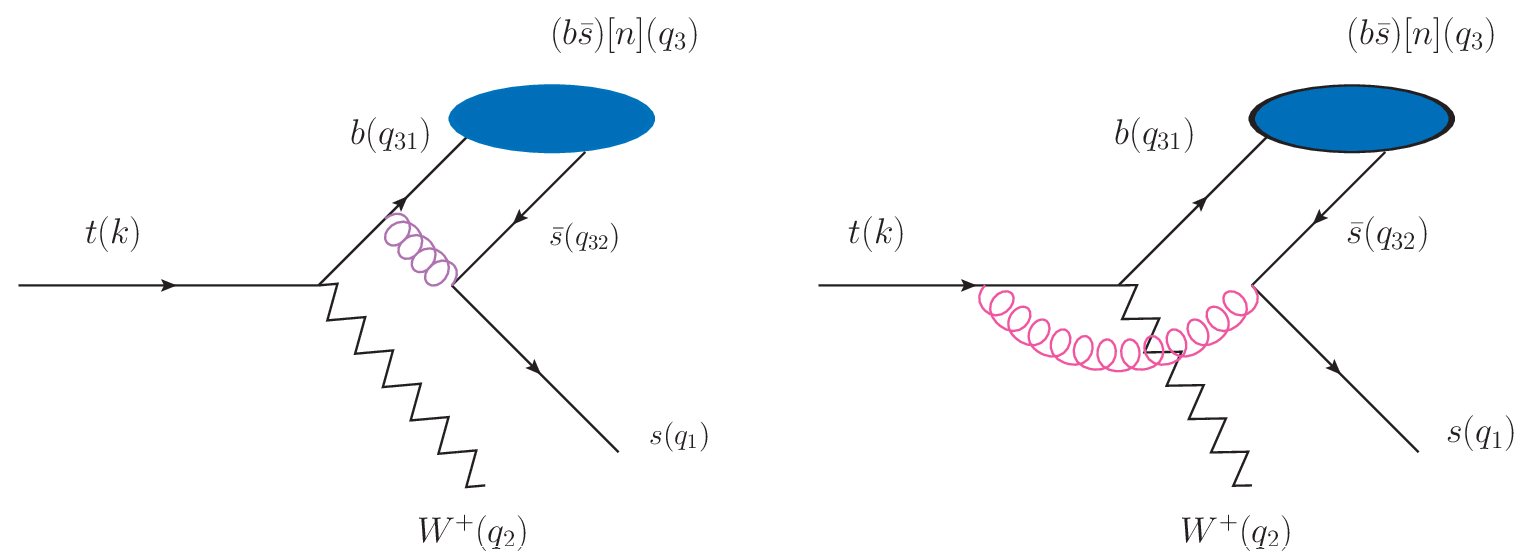}
\caption{(color online). Feynman diagrams for the process $t(k)\rightarrow |(b\bar{s})[n]\rangle(q_3) + W^{+}(q_2)+ s(q_1)$, where $|(b\bar{s})[n]\rangle$ stands for $|(b\bar{s})[^1S_0]\rangle$, $|(b\bar{s})[^3S_1]\rangle$, $|(b\bar{s})[^1P_1]\rangle$, and $|(b\bar{s})[^3P_J]\rangle$ (with $J=[0 , 1, 2]$), respectively.} \label{feyn1}
\end{figure}

\begin{figure}
\includegraphics[width=0.40\textwidth]{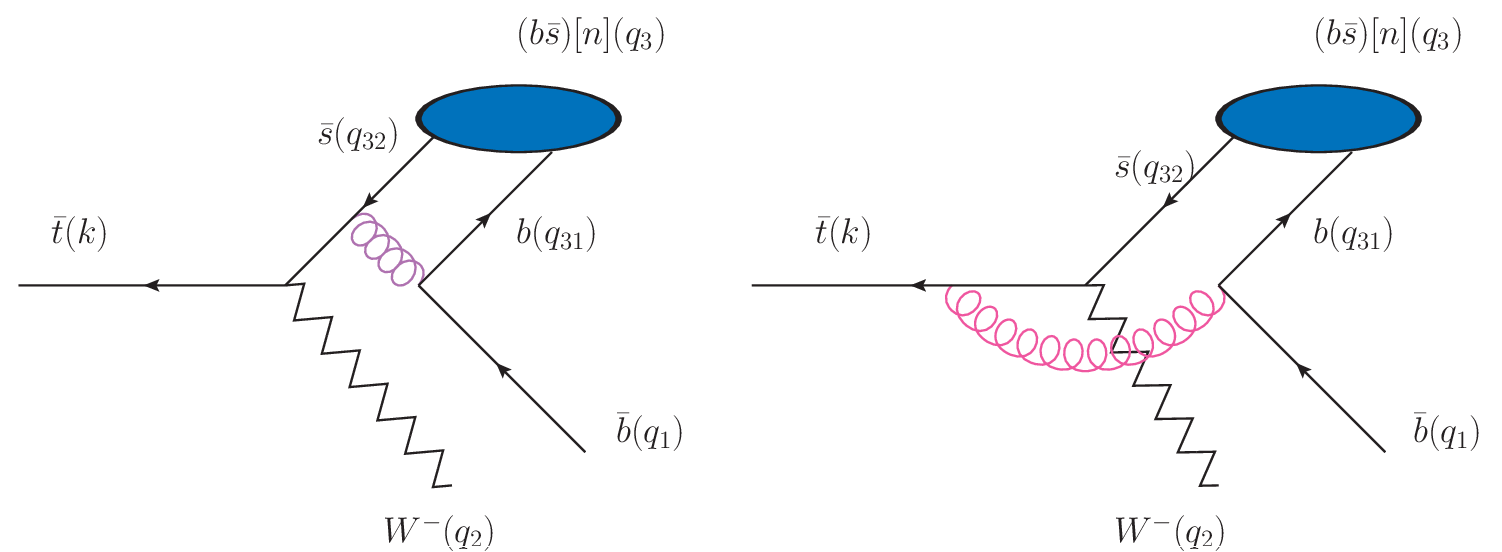}
\caption{(color online). Feynman diagrams for the process $\bar{t}(k)\rightarrow |(b\bar{s})[n]\rangle(q_3) + W^{-}(q_2)+ \bar{b}(q_1)$, where $|(b\bar{s})[n]\rangle$ stands for $|(b\bar{s})[^1S_0]\rangle$, $|(b\bar{s})[^3S_1]\rangle$, $|(b\bar{s})[^1P_1]\rangle$, and $|(b\bar{s})[^3P_J]\rangle$ (with $J=[0 , 1, 2]$), respectively.} \label{feyn2}
\end{figure}
The Feynman diagrams of the two processes, $t(k) \rightarrow |(b\bar{s})[n]\rangle(q_3) + W^{+}(q_2)+s(q_1)$ and $\bar{t}(k) \rightarrow |(b\bar{s})[n]\rangle(q_3) + W^{-}(q_2)+\bar{b}(q_1)$, are presented in Figs.~\ref{feyn1} and \ref{feyn2}, where the intermediate gluon should be hard enough to produce a $s\bar{s}$ pair or $b\bar{b}$ pair, so the amplitudes are pQCD calculable.

These amplitudes of the above-mentioned channels can be generally expressed as
\begin{eqnarray} \label{amplitude}
iM = {\cal{C}_{S}} {\bar {u}_{s i}}({q_1}) \sum\limits_{n = 1}^{m} {{\cal A} _n } {u_{s' j}}({k}),
\end{eqnarray}
where $m$ stands for the number of Feynman diagrams, $s$ and $s'$ are spin states, and $i$ and $j$ are color indices for the outing quark and the initial top quark or antitop quark, respectively. The overall factor ${\cal C}_S=({2gg_s^2 V_{tq}}/{3\sqrt{2}})\delta_{ij}$ stands for the specified quarkonium in the color-singlet, here $V_{tq}$ is the Cabibbo-Kobayashi-Maskawa (CKM) matrix element, where $V_{tq}=V_{tb}$ stands for the processes $t\rightarrow |(b\bar{s})[n]\rangle + W^{+} s $, and $V_{tq}=V_{ts}$ for $\bar{t}\rightarrow |(b\bar{s})[n]\rangle + W^{-}\bar{b}$. While for production the $^{1}S_{0}$ ($^{3}S_{1}$) states of the $(b\bar{s})[n]\rangle$ quarkonium via the decays channel $t(k) \rightarrow |(b\bar{s})[n]\rangle(q_3) + W^{+}(q_2)+s(q_1)$, the amplitudes ${\cal A}_n$ can be written as
\begin{eqnarray}
{\cal A}_1 &=& \left[{\gamma_\alpha} \frac{\Pi^{0(\nu)}_{q_3}(q)}{(q_1 + {q_{32}})^2}{\gamma_\alpha}\frac{(\slashed{q}_1+\slashed{q}_3)+{m_b}}{(q_1+q_3 )^2 - m_b^2} \slashed\epsilon(q_2)({1-\gamma^5}) \right]_{q=0}, \label{A1}\\
{\cal A}_2 &=& \left[{\gamma_\alpha} \frac{\Pi^{0(\nu)}_{q_3}(q)}{(q_1+ {q_{32}})^2}\slashed\epsilon(q_2)({1-\gamma^5}) \frac{(\slashed{q}_2+\slashed{q}_{31}) +{m_t}}{(q_2+q_{31} )^2 - m_t^2} \gamma_\alpha \right]_{q=0}. \label{A2}
\end{eqnarray}

For the $^1P_1$-wave states, ${\cal A}_n$ can be written as
\begin{widetext}
\begin{eqnarray}
{\cal A}^{S=0,L=1}_1 &=& \varepsilon_l^{\mu}(q_3) \frac{d}{dq_\mu} \left[{\gamma_\alpha} \frac{\Pi^{0}_{q_3}(q)}{(q_1 + {q_{32}})^2}{\gamma_\alpha}\frac{(\slashed{q}_1+\slashed{q}_3)+{m_b}}{(q_1+q_3 )^2 - m_b^2} \slashed\epsilon(q_2)({1-\gamma^5})\right]_{q=0}, \label{A3}\\
{\cal A}^{S=0,L=1}_2 &=& \varepsilon_l^{\mu}(q_3) \frac{d}{dq_\mu} \left[{\gamma_\alpha} \frac{\Pi^{0}_{q_3}(q)}{(q_1 + {q_{32}})^2}\slashed\epsilon(q_2)({1-\gamma^5}) \frac{(\slashed{q}_2+\slashed{q}_{31}) +{m_t}}{(q_2+q_{31} )^2 - m_t^2} \gamma_\alpha\right]_{q=0}, \label{A4}
\end{eqnarray}
and the $^3P_J$-wave states ($J=0,1,2$)
\begin{eqnarray}
{\cal A}^{S=1,L=1}_1 &=& \varepsilon^{J}_{\mu\nu}(q_3) \frac{d}{dq_\mu}  \left[{\gamma_\alpha} \frac{\Pi^{\nu}_{q_3}(q)}{(q_1 + {q_{32}})^2}{\gamma_\alpha}\frac{(\slashed{q}_1+\slashed{q}_3)+{m_b}}{(q_1+q_3 )^2 - m_b^2} \slashed\epsilon(q_2)({1-\gamma^5})\right]_{q=0}, \label{A5}\\
{\cal A}^{S=1,L=1}_2 &=& \varepsilon^{J}_{\mu\nu}(q_3) \frac{d}{dq_\mu} \left[{\gamma_\alpha} \frac{\Pi^{\nu}_{q_3}(q)}{(q_1 + {q_{32}})^2}\slashed\epsilon(q_2)({1-\gamma^5}) \frac{(\slashed{q}_2+\slashed{q}_{31}) +{m_t}}{(q_2+q_{31} )^2 - m_t^2} \gamma_\alpha\right]_{q=0}. \label{A6}
\end{eqnarray}
\end{widetext}
Where $m_{b\bar{s}}=m_b+m_s$ is implicitly adopted to ensure the gauge invariance of the hard scattering amplitude. $\varepsilon(q_2)$ is the polarization vector of $W^+$ boson. $\varepsilon_{s}(q_3)$ and $\varepsilon_{l}(q_3)$ are the polarization vectors relating to the spin and the orbit angular momentum of the $|(b\bar{s})[n]\rangle$ quarkonium, $\varepsilon^{J}_{\mu\nu}(q_3)$ is the polarization tensor for the spin triplet $P$-wave states (with $J=[0 , 1, 2]$). The covariant form of the projectors can be conveniently written as
\begin{equation}
\Pi^0_{q_3}(q)=\frac{-\sqrt{m_{b\bar{s}}}}{4{m_b}{m_s}}(\slashed{q}_{32}- m_s) \gamma_5 (\slashed{q}_{31} + m_b)\bigotimes\frac{\mathbf{1_c}}{\sqrt{N_c}},
\end{equation}
and
\begin{equation}
\Pi^\nu_{q_3}(q)=\frac{-\sqrt{m_{b\bar{s}}}}{4{m_b}{m_s}}(\slashed{q}_{32}- m_s) \gamma_\nu (\slashed{q}_{31} + m_b)\bigotimes\frac{\mathbf{1_c}}{\sqrt{N_c}}.
\end{equation}
Here $\mathbf{1_c}$ stands for the unit color matrix with $N_c=3$ for the QCD.

After substituting these projectors into the above amplitudes and doing the possible simplifications, the amplitudes then can be squared, summed over the freedoms in the final state and averaged over the ones in the initial state. The selection of the appropriate total angular momentum quantum number is done by performing the proper polarization sum. If defining
\begin{eqnarray}
\Pi_{\mu\nu}=-g_{\mu\nu}+\frac{q_{3\mu} q_{3\nu}}{M^2}\,,
\end{eqnarray}
the sum over polarization for a spin triplet $S$-state or a spin singlet $P$-state is given by
\begin{eqnarray}
\sum_{J_z}\varepsilon_\mu \varepsilon^*_{\mu'} =\Pi_{\mu\mu'} ,\label{3s1}
\end{eqnarray}
where $J_z=s_z$ or $l_z$ respectively. In the case of $^3P_J$ states, the sum over the polarization is given by \cite{projector}
\begin{eqnarray}\label{3pja}
\varepsilon^{(0)}_{\mu\nu} \varepsilon^{(0)*}_{\mu'\nu'} &=& \frac{1}{3} \Pi_{\mu\nu}\Pi_{\mu'\nu'} \\
\sum_{J_z}\varepsilon^{(1)}_{\mu\nu} \varepsilon^{(1)*}_{\mu'\nu'} &=& \frac{1}{2}
(\Pi_{\mu\mu'}\Pi_{\nu\nu'}- \Pi_{\mu\nu'}\Pi_{\mu'\nu}) \label{3pjb}\\
\sum_{J_z}\varepsilon^{(2)}_{\mu\nu} \varepsilon^{(2)*}_{\mu'\nu'} &=& \frac{1}{2}
(\Pi_{\mu\mu'}\Pi_{\nu\nu'}+ \Pi_{\mu\nu'}\Pi_{\mu'\nu})-\frac{1}{3} \Pi_{\mu\nu}\Pi_{\mu'\nu'} . \label{3pjc}
\end{eqnarray}
The amplitudes ${\cal A} _n$ of the $\bar{t}\rightarrow |(b\bar{s})[n]\rangle+ W^{-}\bar{b}$ in the formulas are similar in the above-mentioned decays channels, so do not listed here to shorten the paper.

Finally, the short-distance decay widths are given by
\begin{eqnarray}
d\hat\Gamma(t\to |(b\bar{s})[n]\rangle+W^{+}s)&=& \frac{1}{2 k^0} \overline{\sum}  |M|^{2} d\Phi_3,\nonumber\\
d\hat\Gamma(\bar{t}\to |(b\bar{s})[n]\rangle+W^{-}\bar{b})&=& \frac{1}{2 k^0} \overline{\sum}  |M|^{2} d\Phi_3.
\end{eqnarray}
where $\overline{\sum}$ means that one needs to average over the spin and color states of the initial particle and to sum over the color and spin of all the final particles.

In the top quark or antitop quark rest frame, the three-particle phase space can be expressed as
\begin{eqnarray}
d{\Phi_3}=(2\pi)^4 \delta^{4}\left(k-\sum_f^3 q_{f}\right)\prod_{f=1}^3 \frac{d^3{\vec{q}_f}}{(2\pi)^3 2~q_f^0}.
\end{eqnarray}

The $1 \to 3$ phase space with massive quark/antiqark in the final state has been calculated to simplify procedures in Refs.~\cite{tbc2,zbc1}. To shorten the paper, we shall not present it here and the interested reader may turn to these two references for the detailed technology. We can not only derive the whole decay widths but also obtain the corresponding differential decay widths that are helpful for experimental studies with the help of the formulas listed in Refs.~\cite{tbc2,zbc1}, such as $d\Gamma/ds_{1}$, $d\Gamma/ds_{2}$, $d\Gamma/d\cos\theta_{12}$, and $d\Gamma/d\cos\theta_{13}$, where $s_{1}=(q_1+q_2)^2$, $s_{2}=(q_1+q_3)^2$, $\theta_{12}$ stands for the angle between $\vec{q}_1$ and $\vec{q}_2$, and $\theta_{13}$ for $\vec{q}_1$ and $\vec{q}_3$.

For the present considered the squared amplitude of top quark semi-exclusive decay process, $t(k) \to (b\bar{s})[n](q_3) + W^{+}(q_2) + s(q_1)$,  there are two massive spinors of top quark and $s$-quark. So to achieve the analytical expression for the squared amplitude becomes too complex and lengthy in the conventional way, especially for $P$-states of $|(b\bar{s})[n]\rangle$ meson. For solved the problem, we adopt `new trace amplitude approach' to deal with the hard scattering amplitude (\ref{amplitude}) for such complicated processes at the amplitude level.  In a difference from the helicity amplitude approach \cite{bcvegpy, helicity1, helicity2, helicity3}, only the coefficients of the basic Lorentz structures are numerical at the amplitude level. However, by using the ``improved trace technology" in Refs.~\cite{tbc2, zbc0, zbc1, zbc2, wbc1, wbc2, lyd, lx}, one can sequentially obtain the squared amplitudes, and the numerical efficiency can also be greatly improved. Detailed processes of the approach can be found in Refs.~\cite{tbc2, zbc0, zbc1, zbc2, wbc1, wbc2}, and here for self-consistency, we shall present its main idea in appendix.

Finally, the partial decay widths over $s_{1}$ and $s_{2}$ can be expressed as
\begin{eqnarray}
\frac{d\Gamma} {ds_{1}ds_{2}}= \frac{\langle{\cal O}^H(n) \rangle}{256 \pi^3 m^3_t}( \overline{\sum}|M|^{2}).
\end{eqnarray}
where $m_t$ is the mass of the top quark.

According to the  CSQCDFF \cite{getg}, the color-singlet nonperturbative matrix element $\langle{\cal O}^H(n) \rangle$ can be related to the Schr${\rm \ddot{o}}$dinger wave function $\psi_{(b\bar{s})}(0)$ at the origin and the first derivative $\psi^\prime_{(b\bar{s})}(0)$ for $S$ and $P$-wave states of the $|(b\bar{s})[n]\rangle$ quarkonium.
\begin{eqnarray}
\langle{\cal O}^H(nS) \rangle &\simeq& |\psi_{\mid(b\bar{s})[nS]\rangle}(0)|^2,\nonumber\\
\langle{\cal O}^H(nP) \rangle &\simeq& |\psi^\prime_{\mid(b\bar{s})[nP]\rangle}(0)|^2.
\end{eqnarray}

As the spin-splitting effects are small, we will not distinguish the difference between the wave function parameters for the spin-singlet $|(b\bar{s})[1^1S_1]\rangle$ and spin-triplet $|(b\bar{s})[1^3S_1]\rangle$ states, also not discriminate the difference among the wave function parameters for $|(b\bar{s})[1^1P_1]\rangle$ and $|(b\bar{s})[1^3P_J]\rangle$ (with $J=[0 , 1, 2]$) in following calculation.

\subsection{Mass of the $|(b\bar{s})[n]\rangle$ quarkonium and matrix element $\langle{\cal O}^H(n)\rangle$}

The mass spectrum of the $|(b\bar{s})[n]\rangle$ bound states has the form
\begin{eqnarray}
M_n(|(b\bar{s})[n]\rangle)=m_b+m_s+E_n(m_b, m_s, V(\vec{r})).
\end{eqnarray}
where $m_b$ and $m_s$ are the current quark masses in the $\overline{MS}$ scheme, and $E_n(m_b, m_s, V(\vec{r}))$ is the energy eigenvalue of the nonrelativistic Schr${\rm \ddot{o}}$dinger equation with a flavor-dependent potential models $V(\vec{r})$~\cite{lx}.

According to the CSQCDFF \cite{getg}, the color-singlet nonperturbative matrix element $\langle{\cal O}^H(n) \rangle$ can be related to the wave function at the origin. In the rest frame of the $|(b\bar{s})[n]\rangle$ quarkonium, it is can be separated the Schr${\rm \ddot{o}}$dinger wave function into radial and angular pieces $\Psi_{nlm}(\vec{r})=R_{nl}(r)Y_{lm}(\theta,\varphi)$, where $n$ is the principal quantum number, and $l$ and $m$ are the orbital angular momentum quantum number and its projection. $R_{nl}(r)$ and $Y_{lm} (\theta,\varphi)$ are the radial wave function and the spherical harmonic function accordingly. The wave function at the origin $\Psi_{|b\bar{s})[nS]\rangle}(0)$ and the first derivative $\Psi'_{|(b\bar{s})[nP]\rangle}(0)$ are related to the radial wave function at the origin   $R_{|(b\bar{s})[nS]\rangle}(0)$ and the first derivative $R^{'}_{|(b\bar{s})[nP]\rangle}(0)$, accordingly.
\begin{eqnarray}
\Psi_{|(b\bar{s})[nS]\rangle}(0)&=&\sqrt{{1}/{4\pi}}R_{|(b\bar{s})[nS]\rangle}(0),\nonumber\\
\Psi'_{|(b\bar{s})[nP]\rangle}(0)&=&\sqrt{{3}/{4\pi}}R'_{|(b\bar{s})[nP]\rangle}(0).
\end{eqnarray}

\section{Numerical results and discussions}

\subsection{Input parameters}

The values of the wave function at the origin for the $|(b\bar{s})[n]\rangle$ quarkonium are related to the number of flavor quark $n_f$, the Regge slope $\alpha'$, the $1$-loop beta-function coefficient $\beta_0$ and the $2$-loop beta-function coefficient $\beta_1$, the renormalization scale parameters ${\Lambda_{\overline{MS}}}$, and so forth, where $\overline{MS}$ is the modified minimal subtraction scheme. In calculating the wave function at the origin of the $|(b\bar{s}')[n]\rangle$ quarkonium within the five potentials \cite{phfp,lx}, we adopt the Regge slope $\alpha'=1.067GeV^{-2}$, and the scale parameters ${\Lambda_{\overline{MS}}}$ as ${\Lambda^{n_f=3}_{\overline{MS}}}$=0.386 GeV, ${\Lambda^{n_f=4}_{\overline{MS}}}$=0.332 GeV, ${\Lambda^{n_f=5}_{\overline{MS}}}$=0.231 GeV \cite{pdg}. The current quark masses, i.e., $m_b=4.19$~GeV and $m_s=0.101$~GeV, are chosen as derived in Ref.~\cite{pdg} for calculating masses of the $|(b\bar{s})[n]\rangle$ quarkonium. The masses of the $|(b\bar{s})[n]\rangle$ quarkonium ($\bar{B}^{0}_{s}$ meson) under the Buchm${\rm \ddot{u}}$ller and Tye potential (B.T. potential)~\cite{lx, pot2, wgs, ec} are presented in Table~\ref{tabrpa}, that can be derived through solving the energy eigenvalue of the Schr${\rm \ddot{o}}$dinger equation~\cite{phfp}. The values of the constituent quark masses $m_b$ and $m_s$ for the $|(b\bar{s})[n]\rangle$ quarkonium and the accordingly quantities $|R_{|(b\bar{s})[nS]\rangle}(0)|^2$, $|R^{'}_{|(b\bar{s})[nP]\rangle}(0)|^2$, and $|R^{''}_{|(b\bar{s})[nD]\rangle}(0)|^2$ are presented in Table~\ref{tabrpb} for the five potential models~\cite{lx}. The five potentials are the B.T. potential, the John L. Richardson potential (J. potential) \cite{lx, jlr}, the K. Igi and S. Ono potential (I.O. potential) \cite{lx, kso, sr}, the  Yu-Qi Chen and Yu-Ping Kuang potential (C.K. potential), and Coulomb-plus-linear potential (the so-called Cor. potential) \cite{lx, pot1, ec, sr}, respectively. During the following calculation, we adopt the values of wave functions at the origin and the constituent quark masses $m_b$ and $m_s$ for the $|(b\bar{s})[n]\rangle$ quarkonium under the B.T. potential in Table \ref{tabrpb} (with $n_{f}=3$) as the central values.

The input other parameters are adopted as the following values \cite{wtd,pdg}: $m_{W}$=80.399 GeV, $m_t = 172.0$ GeV, $|V_{tb}|$=0.88, $|V_{ts}|$=0.0406, $G_F=1.16639\times10^{-5}~{GeV}^{-2}$. We set the renormalization scale to be $m_{Bs}$ of the $|(b\bar{s})\rangle$ quarkonium for leading-order $\alpha_s$ running , which leads to $\alpha_s(m_{B_s})$=0.26. Furthermore, we adopt the constituent quark mass $m_s=0.50$~GeV and $m_b=4.87$~GeV for the $|(b\bar{s})[1S]\rangle$ states, and $m_s=0.69$~GeV and $m_b=5.14$~GeV for the $|(b\bar{s})[1P]\rangle$ states. As the masses of $s$-quark is greater than ${\Lambda_{\overline{MS}}}$, so the above-mentioned processes is pQCD calculable. To ensure the gauge invariance of the hard amplitude, the $|(b\bar{s})[n]\rangle$ quarkonium mass $M$ is set to $m_b+m_{s}$, where $m_{b}$ and $m_{s}$ stand for constituent quark mass.

\subsection{production of the $|(b\bar{s})[n]\rangle$ quarkonium via $t$-quark and $\bar{t}$-quark decays}

As a reference for branching fractions, we calculate the decay widths for the basic processes $t\to b+W^{+}$ and $\bar{t}\to \bar{s}+W^{-}$. Their decay widths can be written as
\begin{eqnarray}
\Gamma &=&\frac {|V_{tq}|^2 G_F |p|} {2 \sqrt{2} \pi {m_t}} (3 {m^2_W} \sqrt{{m_q}^2 +|p|^2}+\nonumber\\
&&2 |p|^2 (~\sqrt {{m^2_q}+|p|^2}+\sqrt{{m^2_W}+|p|^2}~)).
\end{eqnarray}
where $V_{tq}=V_{tb}$ and $m_q=m_b$ are for processes $t\to b+W^{+}$, and $V_{tq}=V_{ts}$ and $m_q=m_s$ for $\bar{t}\to \bar{s}+W^{-}$. $p$ stands for the relative momentum between the final two particles in the rest frame of the top quark or antitop quark.
\begin{eqnarray}
|p| = \frac{\sqrt{(m_t^2 -(m_q -m_W)^2)(m_t^2 -(m_q + m_W)^2)}}{2 m_t}.
\end{eqnarray}

Then, we obtain $\Gamma({t\to b+W^{+}}) = 1.131$~GeV and $\Gamma({\bar{t}\to \bar{s}+W^{-}}) = 2.416$~MeV.

The decay widths and branching fractions for the $|(b\bar{s})[n]\rangle$ quarkonium states through the two decay channels, $t\rightarrow |(b\bar{s})[n]\rangle+W^{+}s$ and $\bar{t}\rightarrow |(b\bar{s})[n]\rangle+W^{-}\bar{b}$, are listed in Tables~ \ref{tabrpc} and~\ref{tabrpd} within the B.T. potential.

\begin{table}
\caption{Mass of the $|(b\bar{s})[n]\rangle$ quarkonium under the B.T. potential \cite{lx, pot2}.}
\begin{tabular}{|c||c|c|c|c|c|c|c|c}
\hline
~~~~~&~$n=1$~&~$n=2$~&~$n=3$~&~$n=4$~&~$n=5$~\\
\hline
$S$~states~({GeV})&5.37&6.08&6.70&7.18&7.60\\
$P$~states~({GeV})&5.83&6.47&7.02&7.53&~\\
$D$~states~({GeV})&6.15&6.75&7.29&~&~\\
$F$~states~({GeV})&6.50&7.06&~&~~&~~\\
\hline\hline
\end{tabular}
\label{tabrpa}
\end{table}

\begin{widetext}
\begin{center}
\begin{table}
\caption{Radial wave functions at the origin and related quantities for the $|(b\bar{s})[n]\rangle$ quarkonium.}
\begin{tabular}{|c||c|c|c|c|c|c|c}
\hline\hline
~~$|(b\bar{s})[n]\rangle$ ~~~&~~~Mass and potential~~~&~~~$n=1$~~~&~~~$n=2$~~~&~~~$n=3$~~~&~~~$n=4$~~~&~~~$n=5$~~~\\
\hline
~&$m_{s}$/$m_{b}$~({GeV})&~0.50/4.87~&~0.73/5.35~&~0.98/5.72~&~1.22/5.96~&~1.46/6.14~\\
~&B.T.($n_{f}$=3) \cite{pot2}&0.6516&0.6737&0.6889&0.7025&0.7808\\
~&B.T.($n_{f}$=4) \cite{pot2}&1.098&0.8043&0.7569&0.8035&0.9311\\
~&B.T.($n_{f}$=5) \cite{pot2}&0.7721&0.7884&0.7484&0.7058&0.8373\\
$S$ states&J.~~~~($n_{f}$=3) \cite{jlr}&0.5505&0.6639&0.8248&0.9730&1.116\\
~~&J.~~~~($n_{f}$=4) \cite{jlr}&0.4814&0.5667&0.6978&0.8185&0.9346\\
~~&J.~~~~($n_{f}$=5) \cite{jlr}&0.3169&0.3559&0.4329&0.5053&0.5760\\
$|R_{|[nS]\rangle}(0)|^2$~~~~~&I.O.~($n_{f}$=3) \cite{kso}&0.2801&0.3497&0.4407&0.5241&0.6041\\
$({GeV}^3)$&I.O.~($n_{f}$=4) \cite{kso}&0.2938&0.3616&0.4536&0.5382&0.6193\\
~&I.O.~($n_{f}$=5) \cite{kso}&0.2996&0.3656&0.4572&0.5417&0.6230\\
~&C.K.($n_{f}$=3) \cite{pot5}&0.3444&0.3911&0.4745&0.5521&0.6276\\
~&C.K.($n_{f}$=4) \cite{pot5}&0.3703&0.4120&0.4958&0.5734&0.6486\\
~&C.K.($n_{f}$=5) \cite{pot5}&0.4042&0.4383&0.5221&0.5997&0.6741\\
~&Cor.~~~~~~~~~~~\cite{pot1}&0.3986&0.5109&0.6608&0.8016&0.9396\\
\hline
~& $m_{s}$/$m_{b}$~({GeV})&~0.69/5.14~&~0.99/5.48~&~1.21/5.81~&~1.40/6.13~&~~\\\
~&B.T.($n_{f}$=3) \cite{pot2}&3.029&17.10&26.21&35.29&~~~\\
~&B.T.($n_{f}$=4) \cite{pot2}&13.32&14.46&37.13&54.23&~~~\\
~&B.T.($n_{f}$=5) \cite{pot2}&8.568&13.04&25.24&39.59&~~~\\
$P$ states&J.~~~~($n_{f}$=3) \cite{jlr}&7.889&22.23&38.81&57.60&~~~\\
~~&J.~~~~($n_{f}$=4) \cite{jlr}&6.036&16.92&29.39&43.46&~~~\\
~~&J.~~~~($n_{f}$=5) \cite{jlr}&2.747&7.324&12.58&20.40&~~~\\
$|R^{'}_{|[nP]\rangle}(0)|^2$~~~~~&I.O.~($n_{f}$=3) \cite{kso} &2.546&7.377&13.11&19.69&~~~\\
$(10^{-2}~{GeV}^5)$&I.O.~($n_{f}$=4) \cite{kso} &2.692&7.764&13.75&20.59&~~~\\
~~&I.O.~($n_{f}$=5) \cite{kso} &2.785&7.974&14.06&20.99&~~\\
~~&C.K.($n_{f}$=3) \cite{pot5}&3.362&9.396&16.22&23.83&~~\\
~~&C.K.($n_{f}$=4) \cite{pot5}&3.632&10.13&17.44&25.56&~~\\
~~&C.K.($n_{f}$=5) \cite{pot5}&3.993&11.09&19.03&27.82&~~\\
~~& Cor.~~~~~~~~~~~\cite{pot1} &3.887&11.59&20.97&31.93&~~\\
\hline
~~&$m_{s}$/$m_{b}$~({GeV})&~0.75/5.40~&~0.98/5.77~&~1.18/6.11~&~~&~~\\
~&B.T.($n_{f}$=3) \cite{pot2}&0.5089&4.272&12.77&~~&~~\\
~&B.T.($n_{f}$=4) \cite{pot2}&1.332&12.93&29.50&~~&~~\\
~&B.T.($n_{f}$=5) \cite{pot2}&1.672&7.200&18.93&~~&~~\\
$D$ states&J.~~~~($n_{f}$=3)~\cite{jlr}&3.628&15.17&37.38&~~&~~\\
~&J.~~~~($n_{f}$=4)~\cite{jlr}&2.384&9.940&24.41&~~&~~\\
~&J.~~~~($n_{f}$=5)~\cite{jlr}&0.5959&2.344&5.443&~~&~~\\
$|R^{''}_{|[nD]\rangle}(0)|^2$~~~~~&I.O.~($n_{f}$=3) \cite{kso} &0.7639&3.257&8.145&~~&~~\\
$(10^{-2}~{GeV}^7)$&I.O.~($n_{f}$=4) \cite{kso} &0.8089&3.443&8.591&~~&~~\\
~&I.O.~($n_{f}$=5) \cite{kso} &0.8430&3.572&8.883&~~&~~\\
~&C.K.($n_{f}$=3) \cite{pot5}&1.023&4.307&10.63&~~&~~\\
~&C.K.($n_{f}$=4) \cite{pot5} &1.101&4.637&11.44&~~&~~\\
~&C.K.($n_{f}$=5) \cite{pot5} &1.206&5.081&12.53&~~&~~\\
~&Cor.~~~~~~~~~~~\cite{pot1}&1.295&5.602&14.16&~~&~~\\
\hline\hline
\end{tabular}
\label{tabrpb}
\end{table}
\end{center}
\end{widetext}

\begin{table}
\caption{Decay widths and branching fractions for the production of the $|(b\bar{s})[n]\rangle$ quarkonium through $t$-quark semiexclusive decays within the B.T. potential ($n_f=3$).}
\begin{tabular}{|c|c|c|c|}
\hline\hline
~$t\rightarrow |(b\bar{s})[n]\rangle+W^{+}s$~&~$\Gamma$(KeV)~&~$\frac{\Gamma{(t\rightarrow |(b\bar{s})[n]\rangle+W^{+}s)}}{{\Gamma{(t\to W^{+}+b)}}}$~\\
\hline
$t\rightarrow |(b\bar{s})[^1S_0]\rangle +W^{+}s$&~~~4528~~~&~~~$4.0\times10^{-3}$\\
\hline
$t\rightarrow |(b\bar{s})[^3S_1]\rangle +W^{+}s$&~~~8614~~~&~~~$7.6\times10^{-3}$\\
\hline
$t\rightarrow |(b\bar{s})[^1P_1]\rangle +W^{+}s$&~~~293.3~~~&~~~$2.6\times10^{-4}$\\
\hline
$t\rightarrow |(b\bar{s})[^3P_0]\rangle +W^{+}s$&~~~119.7~~~&~~~$1.1\times10^{-4}$\\
\hline
$t\rightarrow |(b\bar{s})[^3P_1]\rangle +W^{+}s$&~~~142.5~~~&~~~$1.3\times10^{-4}$\\
\hline
$t\rightarrow |(b\bar{s})[^3P_2]\rangle +W^{+}s$&~~~238.1~~~&~~~$2.1\times10^{-4}$\\
\hline\hline
\end{tabular}
\label{tabrpc}
\end{table}

\begin{table}
\caption{Decay widths and branching fractions for the production of the $|(b\bar{s})[n]\rangle$ quarkonium through $\bar{t}$-quark semiexclusive  decays within the B.T. potential ($n_f=3$).}
\begin{tabular}{|c|c|c|c|}
\hline\hline
~$\bar{t}\rightarrow |(b\bar{s})[n]\rangle+W^{-}\bar{b}$~&~$\Gamma$(eV)~&~$\frac{\Gamma{(\bar{t}\rightarrow |(b\bar{s})[n]\rangle+W^{-}\bar{b})}}{\Gamma{(\bar{t}\to W^{-}+\bar{s})}}$~\\
\hline
$\bar{t}\rightarrow |(b\bar{s})[^1S_0]\rangle +W^{-}\bar{b}$&~~~9.056~~~&~~~$3.7\times10^{-6}$\\
\hline
$\bar{t}\rightarrow |(b\bar{s})[^3S_1]\rangle +W^{-}\bar{b}$&~~~7.095~~~&~~~$2.9\times10^{-6}$\\
\hline
$\bar{t}\rightarrow |(b\bar{s})[^1P_1]\rangle +W^{-}\bar{b}$&~~~0.560~~~&~~~$2.3\times10^{-7}$\\
\hline
$\bar{t}\rightarrow |(b\bar{s})[^3P_0]\rangle +W^{-}\bar{b}$&~~~0.915~~~&~~~$2.4\times10^{-7}$\\
\hline
$\bar{t}\rightarrow |(b\bar{s})[^3P_1]\rangle +W^{-}\bar{b}$&~~~0.531~~~&~~~$2.2\times10^{-7}$\\
\hline
$\bar{t}\rightarrow |(b\bar{s})[^3P_2]\rangle +W^{-}\bar{b}$&~~~0.086~~~&~~~$3.6\times10^{-8}$\\
\hline\hline
\end{tabular}
\label{tabrpd}
\end{table}

\begin{figure}
\includegraphics[width=0.40\textwidth]{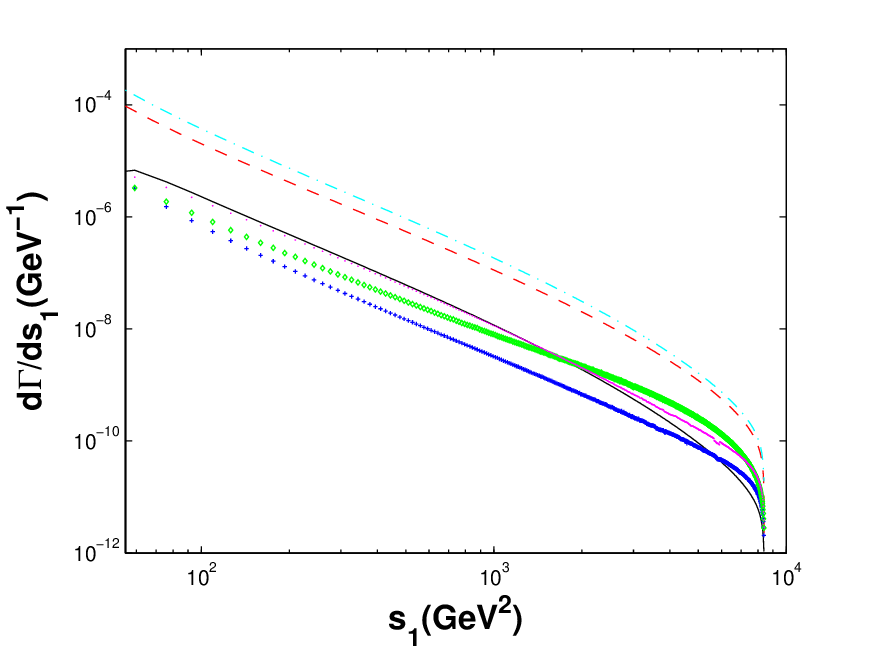}
\includegraphics[width=0.40\textwidth]{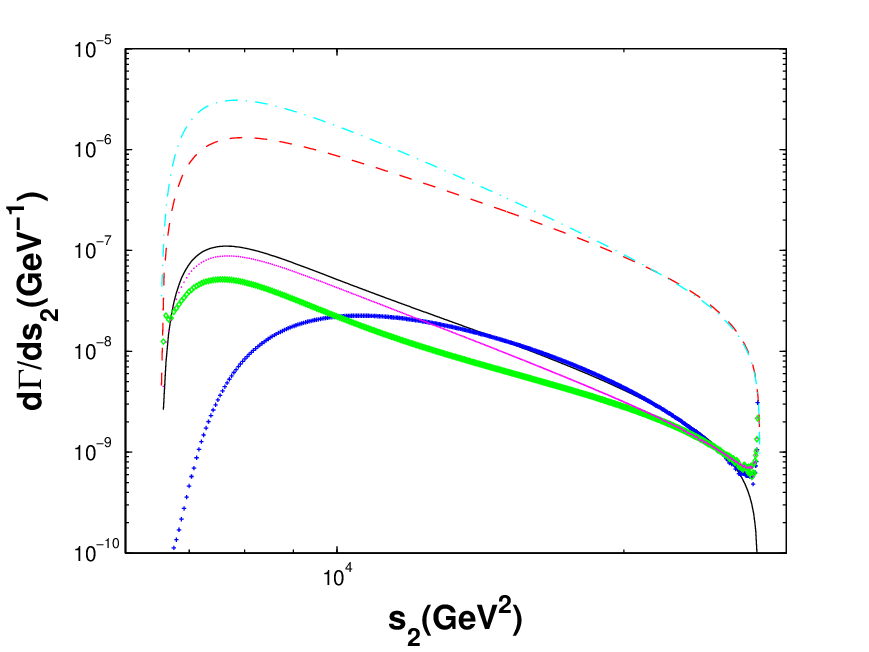}
\caption{(color online). Differential decay widths $d\Gamma/ds_1$ and $d\Gamma/ds_2$ for $ t\rightarrow |(b\bar{s})[n]\rangle +sW^{+}~(n_{f}=3)$, where the dashed line, the dash-dotted line, the solid line, the crosses line, the diamond line, and the dotted line are for $|(b\bar{s})[1^1S_0]\rangle$, $|(b\bar{s})[1^3S_1]\rangle$, $|(b\bar{s})[1^1P_1]\rangle$, $|(b\bar{s})[1^3P_0]\rangle$, $|(b\bar{s})[1^3P_1]\rangle$, and $|(b\bar{s})[1^3P_2]\rangle$, respectively.} \label{tW(bs)sds1ds2}
\end{figure}

\begin{figure}
\includegraphics[width=0.40\textwidth]{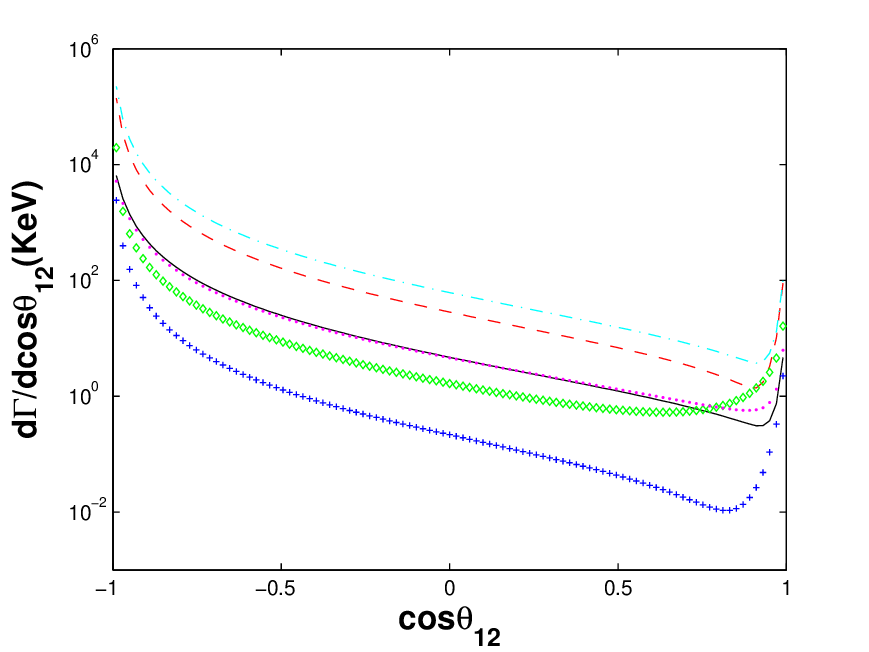}
\includegraphics[width=0.40\textwidth]{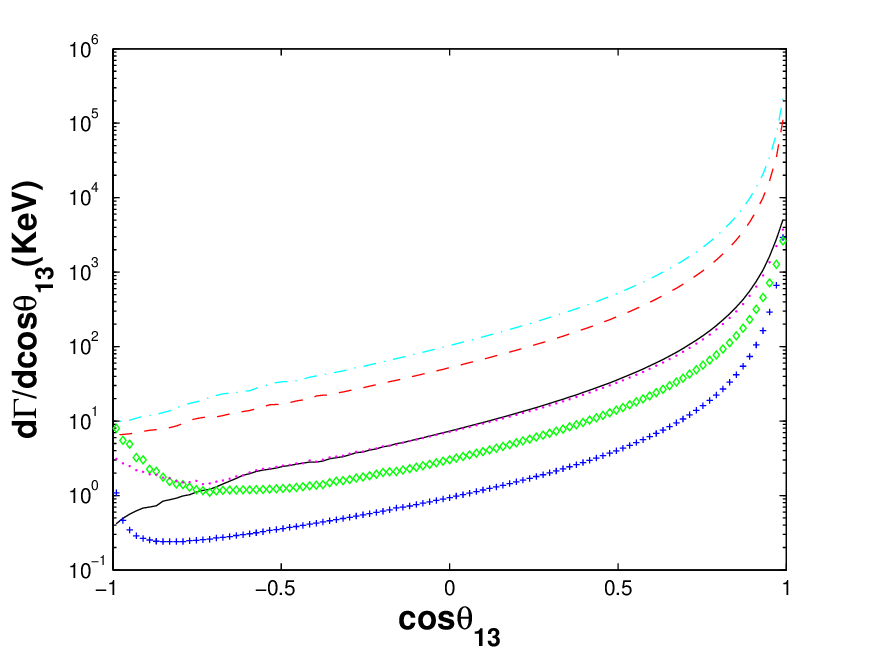}
\caption{(color online). Differential decay widths $d\Gamma/dcos\theta_{12}$ and $d\Gamma/dcos\theta_{13}$ for $ t\rightarrow |(b\bar{s})[n]\rangle +sW^{+}~(n_{f}=3)$, where the dashed line, the dash-dotted line, the solid line, the crosses line, the diamond line, and the dotted line are for $|(b\bar{s})[1^1S_0]\rangle$,  $|(b\bar{s})[1^3S_1]\rangle$, $|(b\bar{s})[1^1P_1]\rangle$, $|(b\bar{s})[1^3P_0]\rangle$, $|(b\bar{s})[1^3P_1]\rangle$, and $|(b\bar{s})[1^3P_2]\rangle$, respectively.} \label{tW(bs)sdcos12cos13}
\end{figure}

\begin{figure}
\includegraphics[width=0.40\textwidth]{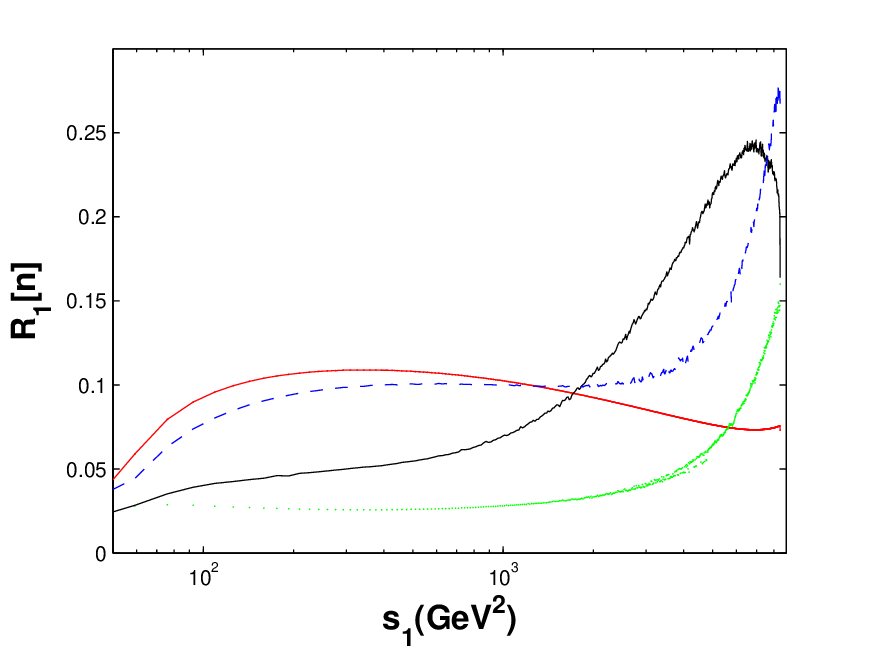}
\includegraphics[width=0.40\textwidth]{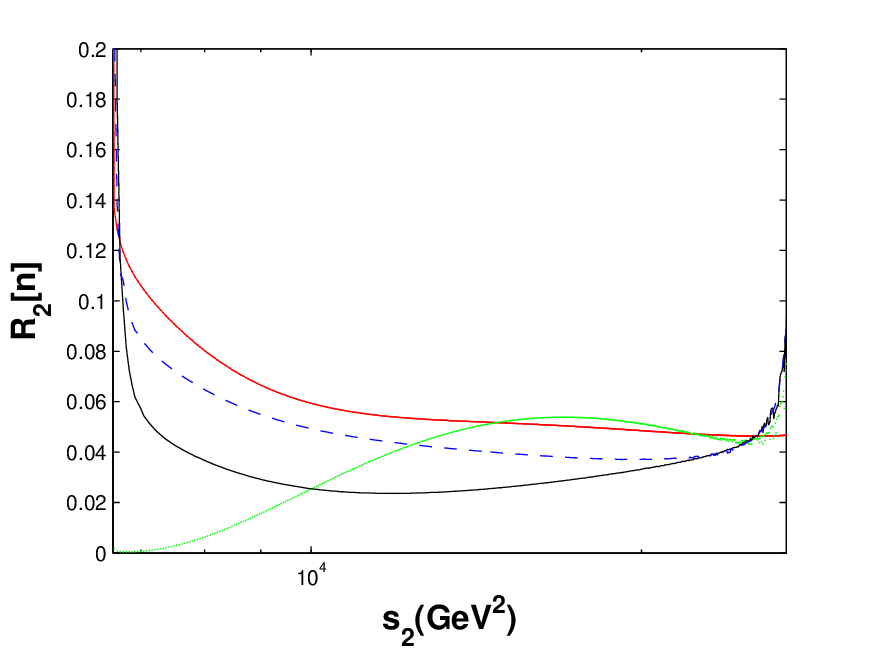}
\caption{(color online). The ratios $R[1]$ and $R[2]$ versus $s_1$ and $s_2$ for the channel $t\rightarrow |(b\bar{s})[n]\rangle+sW^{+}~(n_{f}=3)$. Here the dash-dotted line, the dotted line, the solid line, and the dashed line are for $|(b\bar{s})[1^1P_1]\rangle$, $|(b\bar{s})[1^3P_0]\rangle$, $|(b\bar{s})[1^3P_1]\rangle$, and $|(b\bar{s})[1^3P_2]\rangle$, respectively.} \label{tW(bs)sdr}
\end{figure}

To show the relative importance among different states more clearly, we present the differential distributions
$d\Gamma/ds_{1}$, $d\Gamma/ds_{2}$, $d\Gamma/dcos\theta_{12}$, and $d\Gamma/dcos\theta_{13}$ for the above-mentioned channels in Figs.~\ref{tW(bs)sds1ds2} and~\ref{tW(bs)sdcos12cos13}. Moreover, we define a ratio
\begin{eqnarray}
R_i[n]= \frac{d\Gamma/ds_i(|(b\bar{s})[n]\rangle)}{d\Gamma/ds_i(|(b\bar{s})[1^1S_0]\rangle)}.
\end{eqnarray}
where $i=1, 2$, and $n$ stands for $1^1P_1$, and $1^3P_J$ (with $J=[0, 1, 2]$), respectively. The curves are presented in Fig.~\ref{tW(bs)sdr}. These figures show explicitly that the excited states, i.e., $|(b\bar{s})[1^1P_1]\rangle$ and $|(b\bar{s})[1^3P_J]\rangle$ (with $J=[0 , 1, 2]$), can provide sizable contributions in comparison to the lower state $|(b\bar{s})[1^1S_0]\rangle$ or $|(b\bar{s})[1^3S_1]\rangle$ in almost the entire kinematical region.

For the production of $|(b\bar{s})[n]\rangle$ through channel $t\rightarrow |(b\bar{s})[n]\rangle+sW^{+}$, its total decay width for all the $P$-wave states is $793.6$~KeV, which is about $18\%$ ($9.2\%$) of the decay width of the $|(b\bar{s})[1^1S_0]\rangle$ ($|(b\bar{s})[1^3S_0]\rangle$) meson production. For the $|(b\bar{s})[n]\rangle$ production via $\bar{t}\rightarrow |(b\bar{s})[n]\rangle+\bar{b}W^{-}$, its total decay width for all the $P$-wave states is $2.092$~eV, which is about $29\%$ ($23\%$) of that of $|(b\bar{s})[1^1S_0]\rangle$ ($|(b\bar{s})[1^3S_0]\rangle$). But though $\bar{t}\rightarrow |(b\bar{s})[n]\rangle+\bar{b}W^{-}$ is the CKM matrix element suppressed to $t\rightarrow |(b\bar{s})[n]\rangle+sW^{+}$ by ${|V_{ts}|}^2/{|V_{tb}|}^2\sim0.2\%$, it is enhanced by the phase space, and it is easier to a gluon hadronic generated a $s\bar{s}$ pair than a $b\bar{b}$ pair. So as a combined result, the decay width of $\bar{t}\rightarrow |(b\bar{s})[n]\rangle+\bar{b}W^{-}$ is very smaller than that of $t\rightarrow |(b\bar{s})[n]\rangle+sW^{+}$ by only about $1\times10^{-6}$. Thus, we will do not discuss the uncertainties of the $|(b\bar{s})[n]\rangle$ meson production through $\bar{t}\rightarrow |(b\bar{s})[n]\rangle+\bar{b}W^{-}$ in the next subsection.

Running at the center-of-mass energy $\sqrt{S}=14$ TeV and with luminosity $10^{34} cm^{-2} s^{-1}$ at the LHC, one may expect to produce about $10^8$ $t\bar{t}$-quark per year~\cite{bc1,bc2}. We can estimate the events number of the $|(b\bar{s})\rangle$ quarkonium production through $t$-quark and $\bar{t}$-quark decays, i.e., $4.1\times10^5$ $|(b\bar{s})[1^1S_0]\rangle$ or $|(\bar{b}s)[1^1S_0]\rangle$, $7.8\times10^5$ $|(b\bar{s})[1^3S_1]\rangle$ or $|(\bar{b}s)[1^3S_1]\rangle$, $7.0\times10^4$ $|(b\bar{s})[1P]\rangle$ or $|(\bar{b}s)[1P]\rangle$ events can be obtained per year. It might be possible to find $B^{0}_s$ or $\bar{B}^{0}_s$ through top quark or antitop quark decays, since one may identify these particles through their cascade decay channels, e.g., $B^{0}_s \to D^{-}_s+l^+\nu_l$, $B^{0}_s\to D^{\ast +}_s~D^{-}_s$, $B^{0}_s\to D^{\ast +}_s~D^{\ast -}_s$, or $B^{0}_s\to D^{\pm}_s~K^{\mp}_s$, with clear signals, where $l$ and $\nu_l$ stand for lepton and neutrino accordingly. The possibility of the production of the $|(b\bar{s})[n]\rangle$ or $|(\bar{b}s)[n]\rangle$ quarkonium via $t$-quark or $\bar{t}$-quark decays is worth serious consideration especially when the LHC upgrades~\cite{ab}.

\subsection{Uncertainty analysis}

In the subsection, we discuss the uncertainties for the $|(b\bar{s})[n]\rangle$ quarkonium production through top decays. For the present calculation, their main uncertainty sources include the nonperturbative bound state matrix elements, the CKM matrix elements, the renormalization scale $\mu_R$, and the constituent quark masses $m_b$ and $m_s$. These parameters are the main uncertainty source for estimating the $|(b\bar{s})[n]\rangle$ quarkonium production. Here we only discuss the decay widths of the $|(b\bar{s})[n]\rangle$ quarkonium production through top quark decays under the constituent quark masses of the $|(b\bar{s})[n]\rangle$ quarkonium and the wave functions at the origin of the corresponding constituent quark masses in detail. In the following, we shall concentrate our attention on the uncertainties caused by $m_s$ and $m_b$, whose center values are taken as $m_s=0.50\pm0.10$ GeV and $m_b=4.87\pm0.20$ GeV. And for clarity, when discussing the uncertainty caused by one parameter, the other parameters are fixed to be their center values.

Typical uncertainties for $m_s$, $m_b$, and the values of the wave functions at the origin of the corresponding constituent quark masses are presented in Tables.~\ref{tabrpe} and~ \ref{tabrpf}. In the Tables.~\ref{tabrpe} and~\ref{tabrpf}, it can be found that sizable uncertainties for varying $m_s$ and $m_b$. The decay width will decrease with the increment mass of $m_s$. But the decay width will increase with the increment mass of $m_b$. And such the decay width will slow vary a tendency with a heavier quark mass.
\begin{table}
\caption{Uncertainties for the decay width of the processes $t\to |(b\bar{s})[n]\rangle +W^{+}s$ under the B.T. potential ($n_f=3$).}
\begin{tabular}{|c|c|c|c|c|c|}
\hline\hline
~~&~$m_s$~(GeV)&~0.4~&~0.50~&~0.6~\\
~&~$|R_{|(b\bar{s})[1S]\rangle}|^2~({GeV}^3)$&0.4162&0.6516&0.9316\\
S~states~&$\Gamma(t\rightarrow |(b\bar{s})[1^1S_0]\rangle)$~({MeV})&5.677&4.528&3.730\\
~&$\Gamma(t\rightarrow |(b\bar{s})[1^3S_1]\rangle)$({MeV})&11.41&8.614&6.463\\
\hline
~&~$m_s$~(GeV)&0.59&0.69&0.79\\
~&~$|R'_{|(b\bar{s})[1P]\rangle}|^2~(10^{-2}~{GeV}^5)$&2.022&3.028&4.399\\
P~states&$\Gamma(t\rightarrow |(b\bar{s})[1^1P_1]\rangle)$({KeV})&433.6&293.3&214.7\\
~&$\Gamma(t\rightarrow |(b\bar{s})[1^3P_0]\rangle)$({KeV})&136.6&119.7&109.4\\
~&$\Gamma(t\rightarrow |(b\bar{s})[1^3P_1]\rangle)$({KeV})&203.8&142.5&107.6\\
~&$\Gamma(t\rightarrow |(b\bar{s})[1^3P_2]\rangle)$({KeV})&363.1&238.1&169.0\\
\hline\hline
\end{tabular}
\label{tabrpe}
\end{table}

\begin{table}
\caption{Uncertainties for the decay width of the processes $t\to |(b\bar{s})[n]\rangle +W^{+}s$ under the B.T. potential ($n_f=3$).}
\begin{tabular}{|c|c|c|c|c|c|}
\hline\hline
~&~$m_b$~$(GeV)$~&~4.666~&~4.866~&~5.066~\\
~&$|R_{|(b\bar{s})[1S]\rangle}|^2 $~$({GeV}^3)$&0.6458&0.6516&0.6569\\
S~states&$\Gamma(t\rightarrow |(b\bar{s})[1^1S_0]\rangle)$~({MeV})&4.498&4.528&4.555\\
~&$\Gamma(t\rightarrow |(b\bar{s})[1^3S_1]\rangle)$~({MeV})&8.468&8.614&8.751\\
\hline
~&$m_b$~$(GeV)$~&~4.94~&~5.14~&~5.34~\\
~&~$|R'_{|(b\bar{s})[1P]\rangle}|^2 $~$(10^{-2}~{GeV}^5)$&2.985&3.028&3.070\\
P~states&$\Gamma(t\rightarrow |(b\bar{s})[1^1P_1]\rangle)$({KeV})&290.0&293.3&296.2\\
~&$\Gamma(t\rightarrow |(b\bar{s})[1^3P_0]\rangle)$~({KeV})&125.6&119.7&114.2\\
~&$\Gamma(t\rightarrow |(b\bar{s})[1^3P_1]\rangle)$~({KeV})&141.6&142.5&143.5\\
~&$\Gamma(t\rightarrow |(b\bar{s})[1^3P_2]\rangle)$~({KeV})&233.3&238.1&242.7\\
\hline\hline
\end{tabular}
\label{tabrpf}
\end{table}

\begin{figure}
\includegraphics[width=0.40\textwidth]{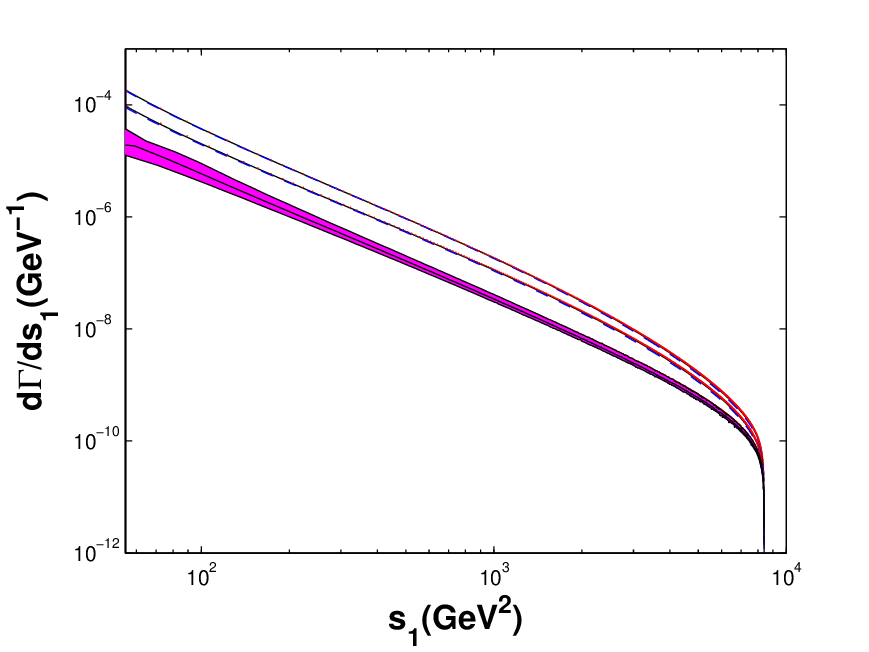}
\includegraphics[width=0.40\textwidth]{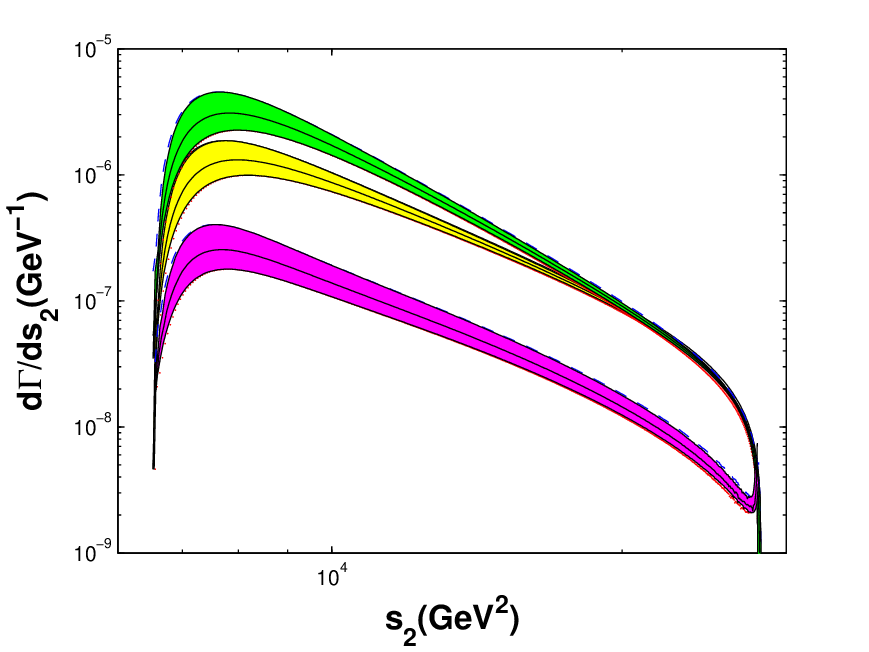}
\caption{(color online). Uncertainties of $d\Gamma/ds_1$ and
$d\Gamma/ds_2$ for $t\to |(b\bar{s})[n]\rangle +sW^{+}$, where contributions from the color-singlet $P$-wave states have been
summed up. The up shaded band, the middle shaded band, the down shaded band are for the uncertainties of $|(b\bar{s})[1^3S_1]\rangle$, $|(b\bar{s})[1^1S_0]\rangle$, and $|(b\bar{s})[1P]\rangle$, respectively.} \label{ds1ds2sum}
\end{figure}

\begin{figure}
\includegraphics[width=0.40\textwidth]{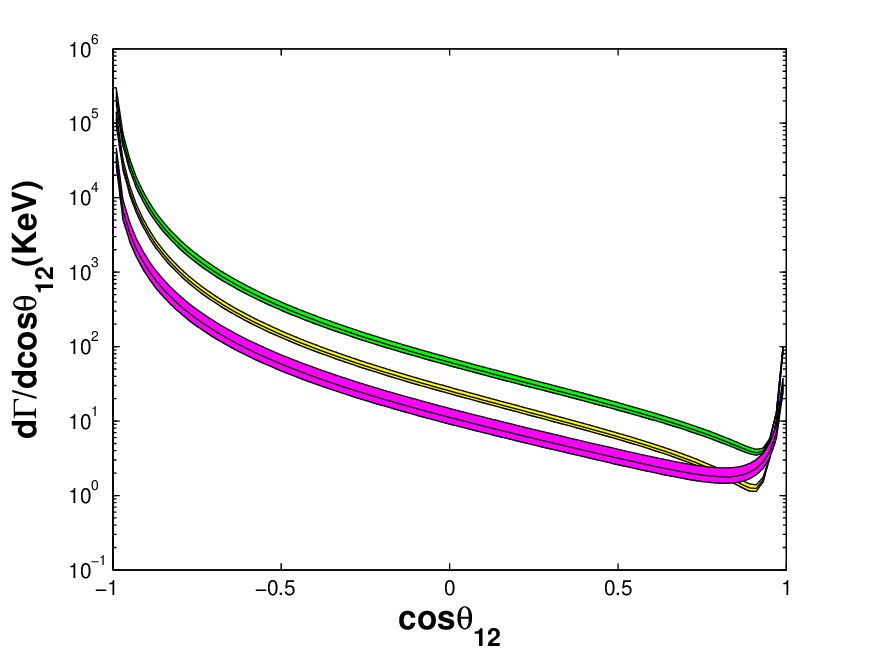}
\includegraphics[width=0.40\textwidth]{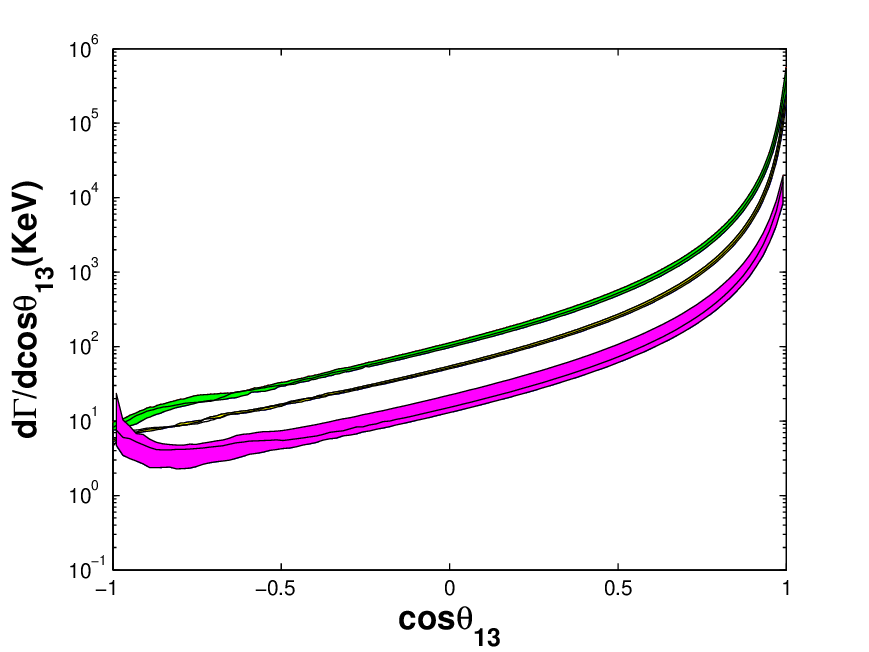}
\caption{(color online). Uncertainties of $d\Gamma/dcos\theta_{12}$ and
$d\Gamma/dcos\theta_{13}$ for $t\to |(b\bar{s})[n]\rangle +sW^{+}$, where contributions
from the color-singlet $P$-wave states have been
summed up. The up shaded band, the middle shaded band, the down shaded band are for the uncertainties of $|(b\bar{s})[1^3S_1]\rangle$, $|(b\bar{s})[1^1S_0]\rangle$ and $|(b\bar{s})[1P]\rangle$, respectively.} \label{dcos12cos13sum}
\end{figure}

The uncertainties are drawn by shaded bands in the Figs.~\ref{ds1ds2sum} and~\ref{dcos12cos13sum}. In the Figs.~\ref{ds1ds2sum} and~\ref{dcos12cos13sum}, the up shaded band is the uncertainties of the $|(b\bar{s})[1^3S_1]\rangle$ meson for the varying constituent quark masses, the middle shaded band is that of the $|(b\bar{s})[1^1S_0]\rangle$ meson, the down shaded band is that of the $|(b\bar{s})[1P]\rangle$. In the above two shaded bands in the Figs.~\ref{ds1ds2sum} and~\ref{dcos12cos13sum}, where the solid-line in the center of the shaded bands is for $m_s=0.50$ GeV and $m_b=4.87$ GeV, the upper edge of the band is for $m_s=0.40$ GeV and $m_b=4.67$ GeV, which the correspondingly radial wave function at the origin is $|R_{|(b\bar{s})[1S]\rangle}|^2 =0.4133~GeV^3$. And the lower edge of the bands is for $m_s=0.60$~ GeV and $m_b=5.07$~GeV, and the correspondingly $|R_{|(b\bar{s})[1S]\rangle}|^2 =0.9404~GeV^3$. In the lowest shaded bands, the solid-line in the center of the shaded bands is for $m_s=0.69$ GeV and $m_b=5.14$ GeV of the $|(b\bar{s})[1P]\rangle$ constituent quark masses, the upper edge of the band is for $m_s=0.59$ GeV and $m_b=4.94$~GeV, and the correspondingly $|R^{'}_{|[1P]\rangle}(0)|^2=1.999\times 10^{-2}~GeV^5$, and the lower edge of the band is for $m_s=0.79$~GeV and $m_b=5.34$~GeV, and $|R^{'}_{|[1P]\rangle}(0)|^2=4.472\times 10^{-2}~GeV^5$, where the contributions of $|(b\bar{s})[1P]\rangle$ form the color-singlet of $|(b\bar{s})[1^1P_1]\rangle$ and $|(b\bar{s})[1^3P_J]\rangle$ (with $J=[0, 1, 2]$) states have been summed up.

Adding all the uncertainties caused by the constituent quark masses $m_s=0.50\pm0.10$ GeV and $m_b=4.87\pm0.20$ GeV in quadrature and the corresponding wave functions at the origin for the process $t\to |(b\bar{s})[n]\rangle +sW^{+}$, we can obtain

\begin{eqnarray}
\Gamma{(t\to |(b\bar{s})[1^1S_0]\rangle +W^{+}s)}&=&4528^{+1149}_{-798.0}\;{\rm KeV},\nonumber\\
\Gamma{(t\to |(b\bar{s})[1^3S_1]\rangle +W^{+}s)}&=&8614^{+2799}_{-2155}\;{\rm KeV},\nonumber\\
\Gamma{(t\to |(b\bar{s})[1P]\rangle +W^{+}s)}~~&=&793.6^{+343.5}_{-192.9}\;{\rm KeV}.
\end{eqnarray}

If the excited $|(b\bar{s})[n]\rangle$ quarkonium states decay to the ground spin-singlet $S$-wave state $|(b\bar{s})[1^1S_0]\rangle$ with $100\%$ efficiency via electromagnetic or hadronic interactions, we can obtain the total decay width of the top quark decay channels under the B.T. potential.
\begin{eqnarray}
\Gamma{(t\to |(b\bar{s})[1^1S_0]\rangle +W^{+}s)}&=&14.19^{+4.36}_{-3.20}\;{\rm MeV} \label{tWbs1}.
\end{eqnarray}

\section{Summary}

In the present paper, for studying the $|(b\bar{s})[n]\rangle$ quarkonium production through $t$-quark or $\bar{t}$-quark decays, we calculate the masses of the $|(b\bar{s})[n]\rangle$ quarkonium under the B.T. potential and the values of the Schr${\rm \ddot{o}}$dinger radial wave function at the origin of the $|(b\bar{s})[n]\rangle$ quarkonium within the five potential models, and made a detailed study on the $|(b\bar{s})[n]\rangle$ quarkonium production via top quark and antitop quark semiexclusive decays channels, $t\to |(b\bar{s})[n]\rangle +W^{+}s$ and $\bar{t}\rightarrow |(b\bar{s})[n]\rangle+W^{-}\bar{b}$, within the CSQCDFF. Results for the $|(b\bar{s})[n]\rangle$ quarkonium states, i.e., $|(b\bar{s})[1^1S_0]\rangle$, $|(b\bar{s})[1^3S_1]\rangle$, $|(b\bar{s})[1^1P_1]\rangle$, and $|(b\bar{s})[1^3P_J]\rangle$ (with $J=[0 ,1 , 2]$) have been presented. And to provide the analytical expressions as simply as possible, we have adopted the ``improved trace technology'' to derive Lorentz-invariant expressions for top quark and antitop quark semiexclusive decay processes at the amplitude level. Such a calculation technology shall be very helpful for dealing with processes with massive spinors.

Numerical results show that excited $|(b\bar{s})[1P]\rangle$ or $|(\bar{b}s)[1P]\rangle$ states in addition to the  ground $1S$-wave states can also provide sizable contributions to the $|(b\bar{s})[n]\rangle$ or $|(b\bar{s})[n]\rangle$ quarkonium production through top quark decays, so one needs to take the excited states into consideration for a sound estimation. If all the excited states decay to the ground state $|(b\bar{s})[1^1S_0]\rangle$, we can obtain the total decay width for the $|(b\bar{s})\rangle$ quarkonium production through top quark decays as shown by Eq.~(\ref{tWbs1}). At the LHC, due to its high collision energy and high luminosity, sizable $|(b\bar{s})[n]\rangle$ or $|(\bar{b}s)[n]\rangle$ quarkonium events can be produced in $t$-quark or $\bar{t}$-quark decays; i.e., about $1.3\times10^6$ $|(b\bar{s})\rangle$ or $|s[n]\rangle$$|(\bar{b}s)\rangle$ quarkonium events per year can be obtained.

\hspace{2cm}

{\bf Acknowledgements}: This work was supported in part by the Natural Science Foundation of China under Grant No.11347024, the Scientific and Technological Research Program of Chongqing Municipal Education Commission under Grant No. KJ1401313 and the Research Foundation of Chongqing University of Science and Technology under Grant No. CK2016Z03, the Natural Science Foundation Project of CQCSTC under Grant No. 2014jcyjA00030.

\end{document}